# Design, integration, and test of the scientific payloads on-board the HERMES constellation and the SpIRIT mission

Y. Evangelista[*a,b], F. Fiore[c], R. Campana[d,e], F. Ceraudo[a], G. Della Casa[f], E. Demenev[g], G. Dilillo[a], M. Fiorini[h], M. Grassi[i], A. Guzman[j], P. Hedderman[j], E. J. Marchesini[d], G. Morgante[d], F. Mele[k], P. Nogara[l], A. Nuti[a], R. Piazzolla[m], S. Pliego Caballero[j], I. Rashevskaya[n], F. Russo[l], G. Sottile[l], C. Labanti[d], G. Baroni[d], P. Bellutti[g], G. Bertuccio[k], J. Cao[o], T. Chen[o], I. Dedolli[k], M. Feroci[a,b], F. Fuschino[d], M. Gandola[g], N. Gao[o], F. Ficorella[g], P. Malcovati[i], A. Picciotto[g], A. Rachevski[p], A. Santangelo[j], C. Tenzer[j], A. Vacchi[f], L. Wang[o], Y. Xu[o], G. Zampa[p], N. Zampa[f,q] and N. Zorzi[g], on behalf of the HERMES collaboration

[a]INAF-IAPS Rome, Italy; [b]INFN sez. Roma Tor Vergata, Italy; [c]INAF-OATS, Trieste, Italy; [d]INAF-OAS Bologna, Italy; [e]INFN sez. Bologna, Italy; [f]Università di Udine; [g]Fondazione Bruno Kessler, Trento, Italy; [h]INAF-IASF Milano, Italy; [i]University of Pavia, Pavia, Italy; [j]IAAT University of Tübingen, Germany; [k]Politecnico di Milano, Como, Italy; [l]INAF-IASF Palermo, Italy; [m]Agenzia Spaziale Italiana, Rome, Italy; [n]TIFPA-INFN, Trento, Italy; [o]Institute of High Energy Physics, Chinese Academy of Sciences, Beijing, China; [p]INFN sez. Trieste, Italy; [q]INFN sez. Udine, Italy.

## ABSTRACT

HERMES (High Energy Rapid Modular Ensemble of Satellites) is a space-borne mission based on a constellation of nano-satellites flying in a low-Earth orbit (LEO). The six 3U CubeSat buses host new miniaturized instruments hosting a hybrid Silicon Drift Detector/GAGG:Ce scintillator photodetector system sensitive to X-rays and gamma-rays. HERMES will probe the temporal emission of bright high-energy transients such as Gamma-Ray Bursts (GRBs), ensuring a fast transient localization (with arcmin-level accuracy) in a field of view of several steradians exploiting the triangulation technique. With a foreseen launch date in late 2023, HERMES transient monitoring represents a keystone capability to complement the next generation of gravitational wave experiments. Moreover, the HERMES constellation will operate in conjunction with the Space Industry Responsive Intelligent Thermal (SpIRIT) 6U CubeSat, to be launched in early 2023. SpIRIT is an Australian-Italian mission for high-energy astrophysics that will carry in a Sun-synchronous orbit (SSO) an actively cooled HERMES detector system payload. On behalf of the HERMES collaboration, in this paper we will illustrate the HERMES and SpIRIT payload design, integration and tests, highlighting the technical solutions adopted to allow a wide-energy-band and sensitive X-ray and gamma-ray detector to be accommodated in a 1U Cubesat volume.

## 1. INTRODUCTION

The most dramatic events in the Universe, the death of stars and the coalescence of compact objects to form a new black hole, produce among the most luminous objects in the Universe: Gamma Ray Bursts (GRB). However, most of the light is produced quite far from where the action is, i.e., the newborn event horizon, the accretion disk, and the region from which a relativistic jet is launched. On the other hand, gravitational waves (GWs), encoding the rapid/relativistic motion of compact objects, give us a direct look into the innermost regions of these systems, providing precise information on space-time dynamics such as mass, spin, inclination, and distance. This information can be greatly enhanced by identifying the context in which the event occurs, which can be done via electromagnetic observations, as the

[*]Send correspondence to Y. Evangelista: yuri.evangelista@inaf.it; https://www.hermes-sp.eu

GW/GRB170817 event strikingly showed [1]. While this event hinted at the enormous potential of multi-messenger astrophysics, it has so far remained unique, preventing the full impact of the multi-messenger approach from occurring.

The operation of an efficient X-ray all-sky monitor with good localization capabilities will have a pivotal role in bringing multi-messenger astrophysics to maturity and will fully exploit the huge advantages provided by adding a further dimension to our capability to investigate cosmic sources. The HERMES (*High Energy Rapid Modular Ensemble of Satellites*) program offers a fast-track and affordable complement to more complex and ambitious missions for relatively bright events.

HERMES [2] is a mission concept based on a constellation of nano-satellites (3U CubeSats) in low Earth orbit. The constellation is built upon a twin project: the HERMES Technological Pathfinder (HERMES-TP), funded by the Italian Ministry for education, university and research and the Italian Space Agency, and the HERMES Scientific Pathfinder (HERMES-SP), funded by the European Union's Horizon 2020 Research and Innovation Programme under Grant Agreement No. 821896. Both projects (HERMES-TP and HERMES-SP) provide three complete satellites (payload and service module) to the constellation, aiming at demonstrating that fast GRB detection and localization is feasible with disruptive technologies on-board miniaturized spacecrafts, mostly exploiting commercial off-the shelf (COTS) components. Moreover, the Italian Space Agency approved and funded the participation to the SpIRIT (*Space Industry – Responsive – Intelligent – Thermal Nanosatellite*) CubeSat. The SpIRIT project, which is supported by the Australian Space Agency and led by University of Melbourne, will host a HERMES-like detector, thus providing a seventh unit to the HERMES constellation [2].

At the core of the HERMES-Pathfinder mission is a hybrid detector concept that is capable of measuring both soft X-rays as well as $\gamma$-rays. The selected design consists of GAGG:Ce scintillator crystals optically coupled to Silicon Drift Detectors (SDDs). The SDDs can directly detect the soft X-rays up to ∼30 keV, while for higher energies the SDDs collect the light produced by the $\gamma$-rays in the scintillator crystals, effectively extending the sensitivity to the MeV range.

## 2. HERMES PAYLOAD REQUIREMENTS

The HERMES Pathfinder Scientific Requirements flow-down to ambitious payload requirements, including broad energy band, high detection efficiency, good energy resolution, sub-microsecond temporal resolution, compact and lightweight design, reliable operation in a quite broad range of space environments (e.g., temperature, radiation damage, etc.).

To fulfill the broad energy range requirement, an integrated detector [4] has been developed to exploit a "double detection" mechanism, with a partial overlap of the two different detection systems efficiency around ~30 keV. Detection of soft X-rays (hereafter *X-mode*) is obtained by a segmented solid state detector employing custom designed Silicon Drift Detectors (SDD) [5], with a cell size of about 0.45 $cm^2$, which allows to attain a low noise level (of the order of a few tens of $e^-$ rms at room temperature) and correspondingly a low energy threshold for the detection of X-ray radiation (~3 keV). On the other hand, detection of hard X-rays and γ-rays (hereafter *S-mode*) is obtained exploiting the conversion of high-energy photons into visible light by means of scintillator crystals. Scintillation generated optical photons are then collected and converted into electric charge in the same photodetector (SDD) used to directly detect X-ray photons. In this case, the SDDs act as a photodiode and produces an amplitude charge signal proportional to the amount of scintillator light collected.

The discrimination between the two signals in the SDD (soft X-rays or optical photons) is achieved exploiting a segmented detector design. Each scintillator crystal is read-out by two SDD cells, so that events detected in only one SDD are associated to soft x-rays converted in a single SDD cell, while events detected simultaneously in more than one SDD are generated by the optical light produced in the scintillator by an incoming hard X-ray/γ-ray. Table 1 summarizes the main HERMES Pathfinder Payload Requirements.

Table 1 HERMES Pathfinder Payload requirements

| Requirement | Condition | Value |
|---|---|---|
| **Sensitivity** | $E \leq 20$ keV (GRB short/long) | $\leq 2$ photons/s/cm$^2$ |
| | $50 \leq E \leq 300$ keV (GRB short/long) | $\leq 1$ photons/s/cm$^2$ |
| **Energy band** | $E_{low} \leq 5$ keV, $E_{high} \geq 500$ keV | |
| **Peak effective area** | X-mode | $\geq 50$ cm$^2$ |
| | S-mode | $\geq 50$ cm$^2$ |
| **Lower energy threshold** | | $\leq 5$ keV |
| **Energy resolution EOL** | between 5.0 and 6.0 keV | $\leq 1$ keV FWHM |
| | between 50.0 and 60.0 keV | $\leq 5$ keV FWHM |
| **Time resolution (1σ)** | X-mode | $\leq 400$ ns |
| | S-mode | $\leq 250$ ns |
| **Time accuracy (1σ)** | GPS locked | $\leq 100$ ns |
| | GPS unlocked (up to 1500 s) | $\leq 200$ ns |
| **Field of view** | | $\geq 3$ sr FWHM |
| **Background rate 50–300 keV** | | $\leq 1.5$ counts/s/cm$^2$ |
| **Background rate 20–300 keV** | | $\leq 12$ counts/s/cm$^2$ |
| **Background knowledge** | | $\leq 5\%$ |
| **Maximum sustainable GRB flux** | | 40000 counts/s |
| **On-board memory** | | $\geq 16$ Gbit |
| **Mass allocation** | | $< 1.8$ kg |
| **Volume allocation** | | $\leq 10 \times 10 \times 12.5$ cm$^3$ (1.25 U) |
| **Power allocation** | | $\leq 5$W |
| **Detector operative T range** | | –30 °C ÷ +10 °C |
| **P/L non operative T range** | | –40 °C ÷ +80 °C |

## 3. HERMES PAYLOAD DESIGN

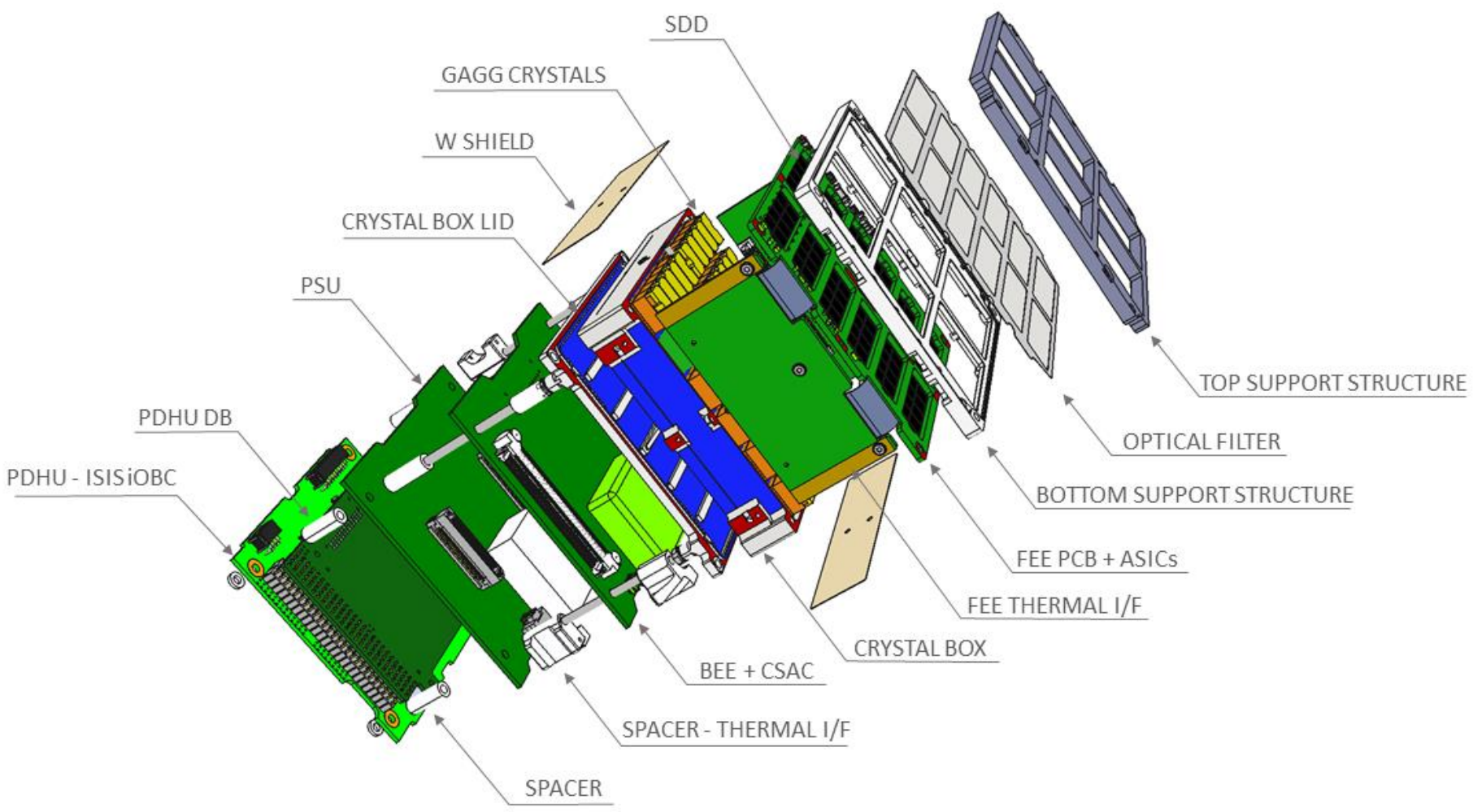

Figure 1 Exploded view of the HERMES Pathfinder Payload

Figure 1 shows an exploded view of the HERMES Pathfinder payload. The payload is composed by 4 main subsystems: the detector assembly (DA), the back-end electronic board (BEE), the power-supply electronic board (PSU) and the payload data handling unit (PDHU). In the following we provide detailed description of each payload subsystem.

### 3.1 Detector Assembly (DA)

An exploded view of the HERMES Pathfinder detector assembly is shown in Figure 1.

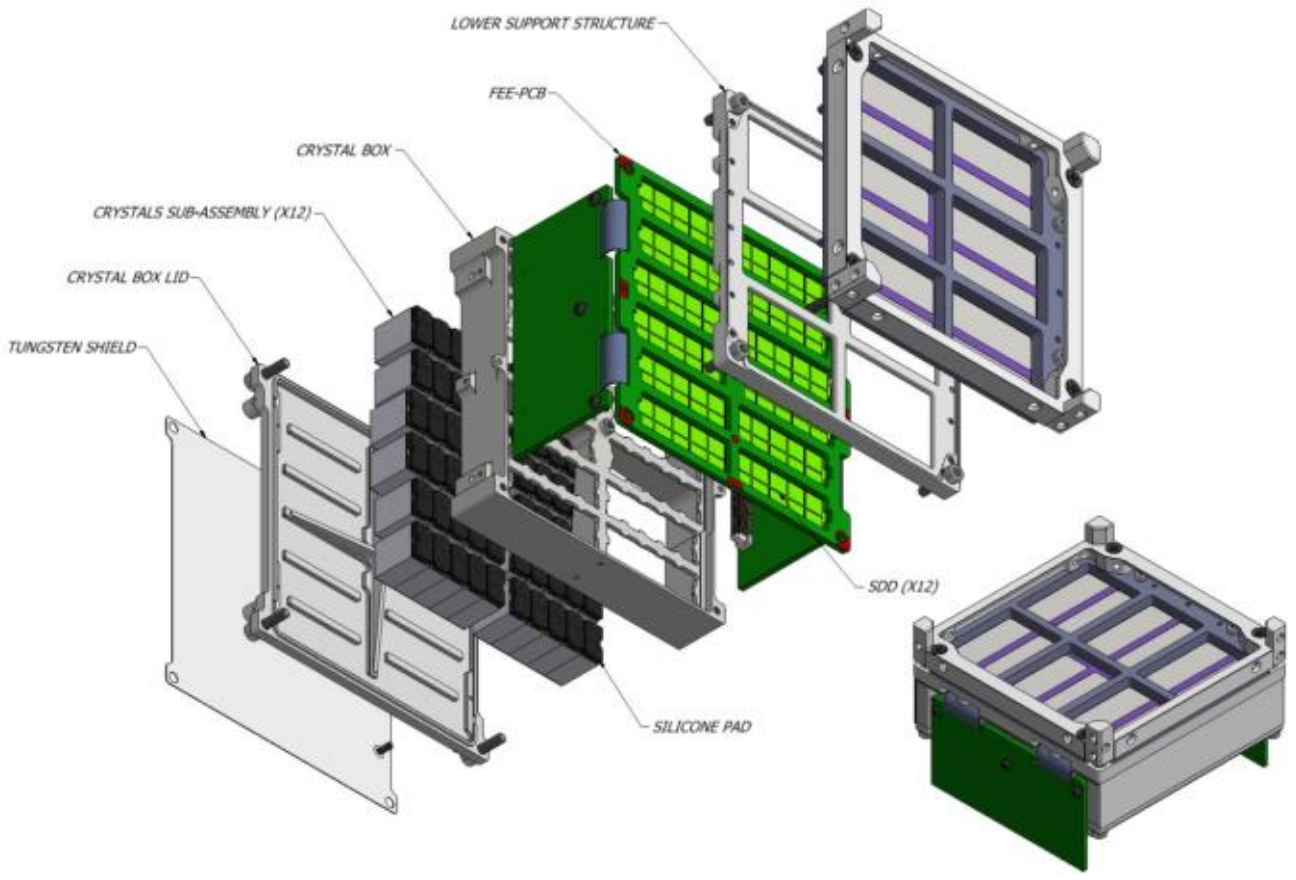

Figure 2 Exploded view of the HERMES Pathfinder and SpIRIT Detector Assembly (DA)

The main components of the HERMES Pathfinder and SpIRIT Detector Assembly are:

- **Optical filter**: made from 300 nm aluminum deposited on a thin (1μm) polyimide foil. The optical filter prime task is to prevent O/UV light from reaching the SDD detector, minimizing the current noise generated in the NIR/O/UV band. In addition, being the HERMES P/L thermal design based on passive cooling only, the filter also contributes to the overall thermal design of the detector assembly. The filter has been developed and manufactured by the Institute of High Energy Physics (IHEP) of Beijing, China.

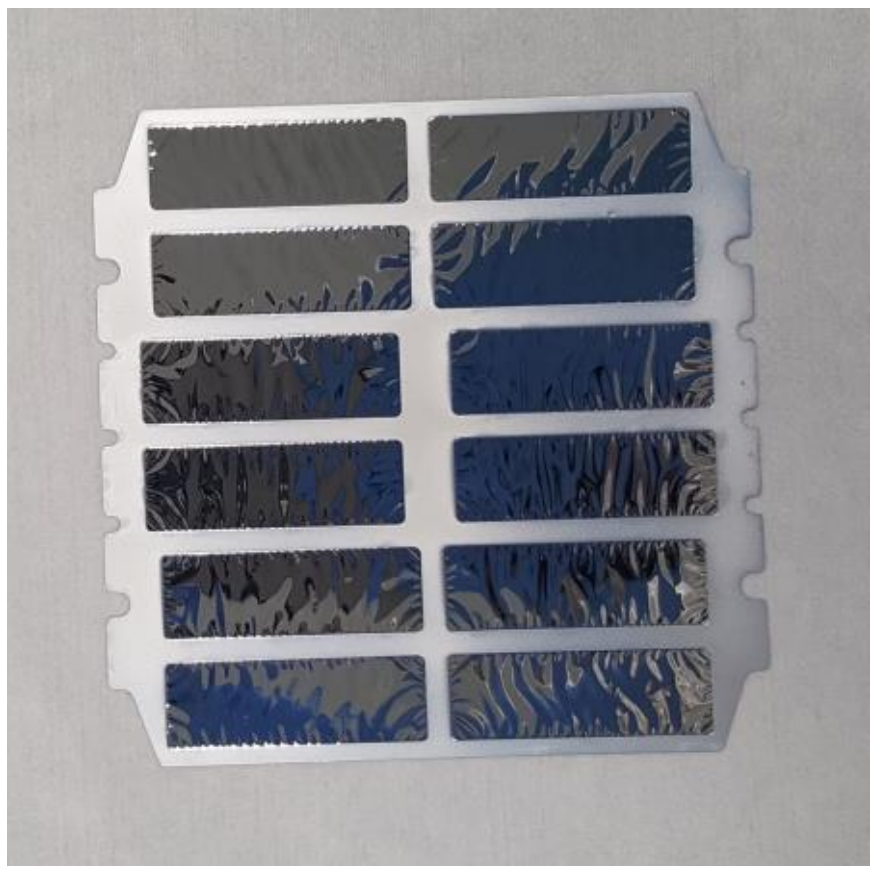

Figure 3 HERMES Optical Filter

- **Detector support structure**: made of passivated AISI-316 stainless steel, the detector support structure is composed of a bottom support structure and a top support structure with the optical filter placed in the middle. This assembly represents the mechanical and thermal interface with the payload top rib. The bottom support structure has an external frame matching the outline of the FEE-PCB, allowing for the required light-tightness.

- **Multilayer insulation**: An additional multilayer insulation film (MLI) will be mounted on top of the detector after the final mechanical assembly. This is to ensure a suitable operating temperature for the SDDs in the foreseen orbital environment.
- The **HERMES SDDs** (Figure 4) are based on the state-of-the-art results achieved within the framework of the Italian ReDSoX collaboration[1], with the combined design and manufacturing expertise of INFN-Trieste and Fondazione Bruno Kessler (FBK, Trento). Each SDD array (Figure 4) is composed by 2×5 cells on a 450 μm thick silicon substrate, with a cell of 7.44×6.05 $mm^2$ for a total of 4.5 $cm^2$ sensitive area. The whole matrix is surrounded by a 1.2 mm wide guard region which allows for a correct electric field termination. The overall geometric area of the SDD array is 39.6×14.5 $mm^2$. By exploiting the single-side biasing technique [6], all the SDD bonding pads are placed on the detector *n*-side (i.e., the anode side), thus providing a homogenous light entrance window for scintillation light readout. Four 500 μm wide metallization strips are present on the *p*-side (optical window side) to ensure negligible optical crosstalk of the scintillation light emitted by adjacent crystals

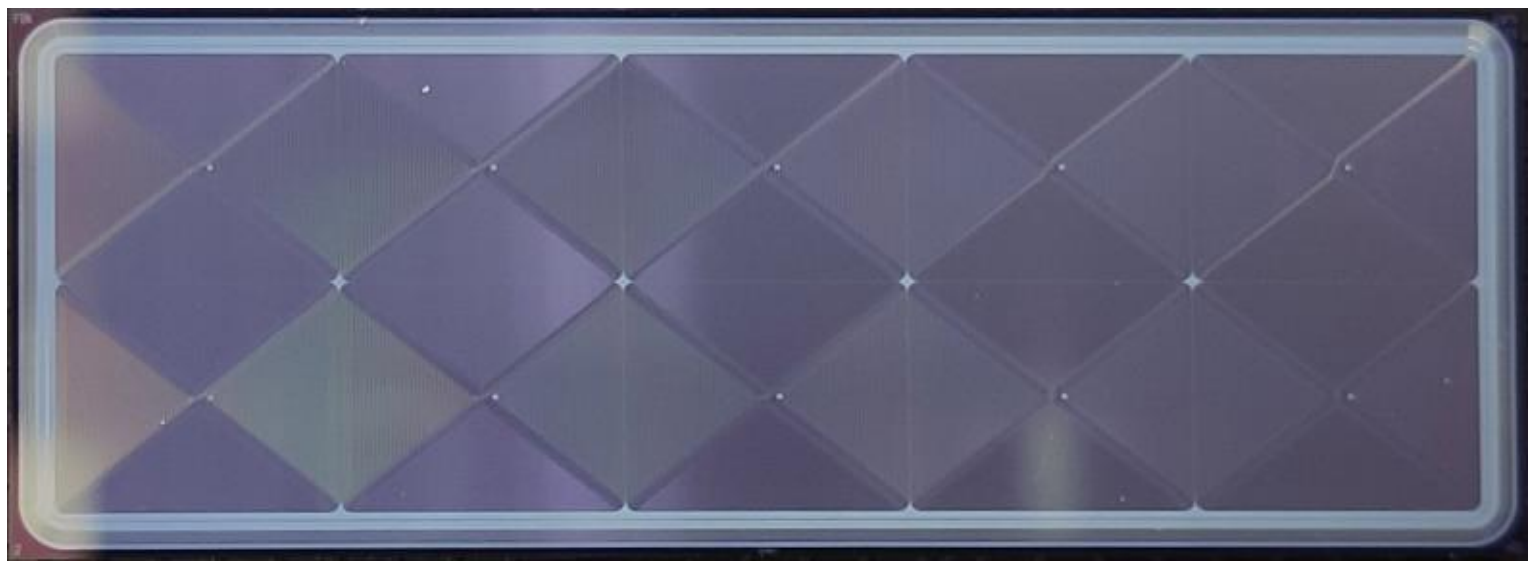

Figure 4 2×5 HERMES SDD matrix (n-side/anode side). Each SDD cell is 7.44×6.05 $mm^2$.

- **Crystal scintillators**: The HERMES scintillators are cerium-doped gadolinium-aluminum-gallium garnet crystals ($Gd_3Al_2Ga_3O_{12}$ or Ce:GAGG), developed firstly in Japan around 2010, and commercially available since 2014 [7][8]. Ce:GAGG scintillators have high intrinsic light output (~50000 photons/MeV), no intrinsic background, no hygroscopicity, fast radiation decay time of ~90 ns, high density (6.63 $g/cm^3$), peak light emission at 520 nm and high effective mean atomic number (54.4). All these characteristics make the Ce:GAGG scintillators the optimal choice for the HERMES detector. The crystals have a 6.94×12.10 $mm^2$ cross section and a 15 mm thickness, with chemically polished faces. Each crystal is optically coupled to two adjacent SDD cells by means of a 3.3 mm thick Polydimethylsiloxane encapsulant (DOWSIL 93-500), allowing for a reliable discrimination of X-ray and γ-ray photons via multiplicity analysis. The crystals are individually wrapped with a 3M DF2000MA specular reflector film and grouped in sub-assemblies of five crystals, with one sub-assembly coupled to an individual SDD array (Figure 5). With the reported optical configuration, the crystal measured an effective light output at room temperature is ~28 $e^-$/keV, which translates in ~14 $e^-$/keV collected by each SDD channel reading-out the scintillator.

[1] http://redsox.iasfbo.inaf.it

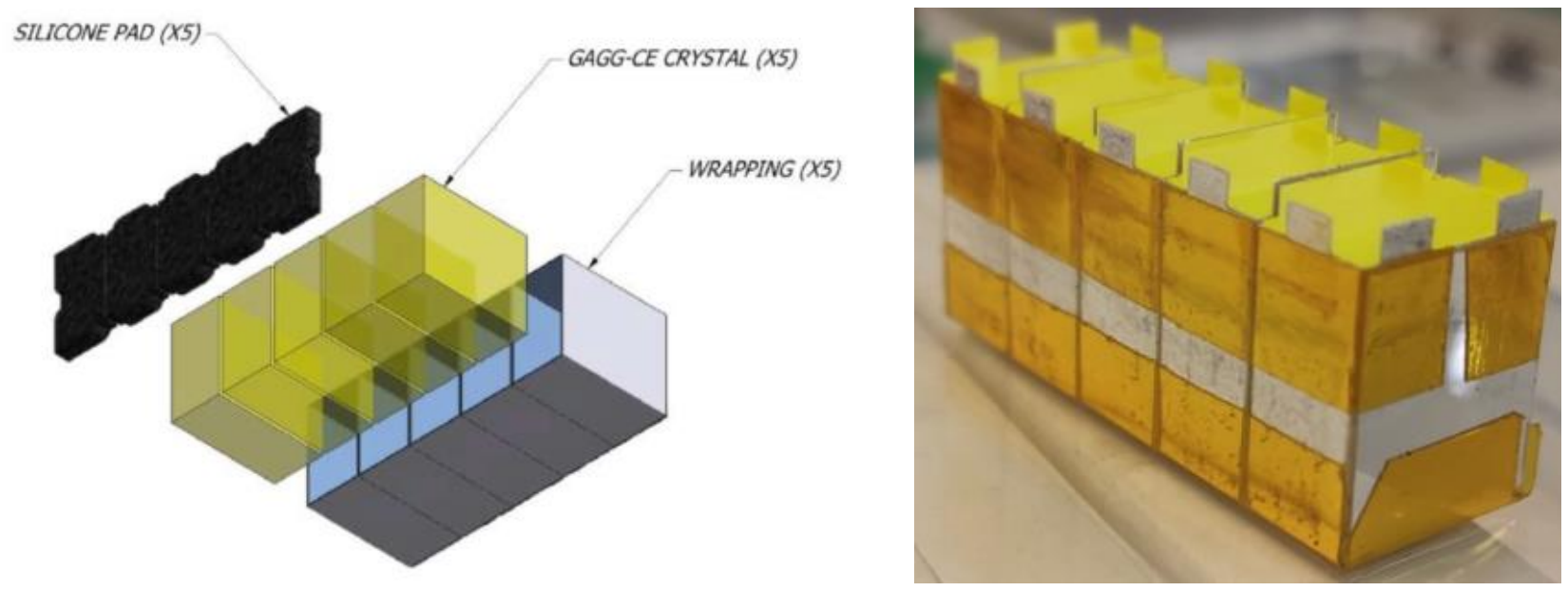

Figure 5 Ce:GAGG crystal optical assembly

- **Front-end electronics**: The Front-end electronic board (FEE, left panel of Figure 6) accommodates the 12 2×5 SDD matrices, 120 LYRA-FE ASICs, 4 LYRA-BE ASICs [9][12], and ancillary passive electronic components. The PCB is manufactured with a rigid-flex technology, allowing the 90° bending of the two side wings. Each wing hosts 2 LYRA-BE ASICs and a board-to-board connector providing the FEE electrical interface with the back-end electronic board. The 120 LYRA-FE chips are mounted on the top PCB layer, while the 12 SDD arrays are glued on the bottom PCB layer by means of DOWSIL 3145 silicone adhesive. Aluminum wire bondings provide the electrical connection between the ASICs and the FEE PCB (chip-to-board), the SDD biases (chip-to-board), and connection between the SDD anodes and the LYRA-FE input pads (chip-to-chip). The 0.35-µm CMOS read-out ASIC design (Figure 6, right panel) is based on the heritage of the VEGA ASIC [11][12], developed by Politecnico of Milano and University of Pavia. To optimize the system noise performance, a distributed architecture has been adopted for the HERMES ASICs. The Front-End ICs (LYRA-FE), which include the preamplifier, the first shaping stage and signal line-transmitter, are placed as close as possible to each SDD anode. The current signal produced by the first-stage pulse shaper is transferred to one-channel of the back-end IC (LYRA-BE), where it is collected by the current receiver block and further processed. Each LYRA-BE can manage the signals produced by 30 LYRA-FE ICs, corresponding to one fourth of the HERMES detection plane.

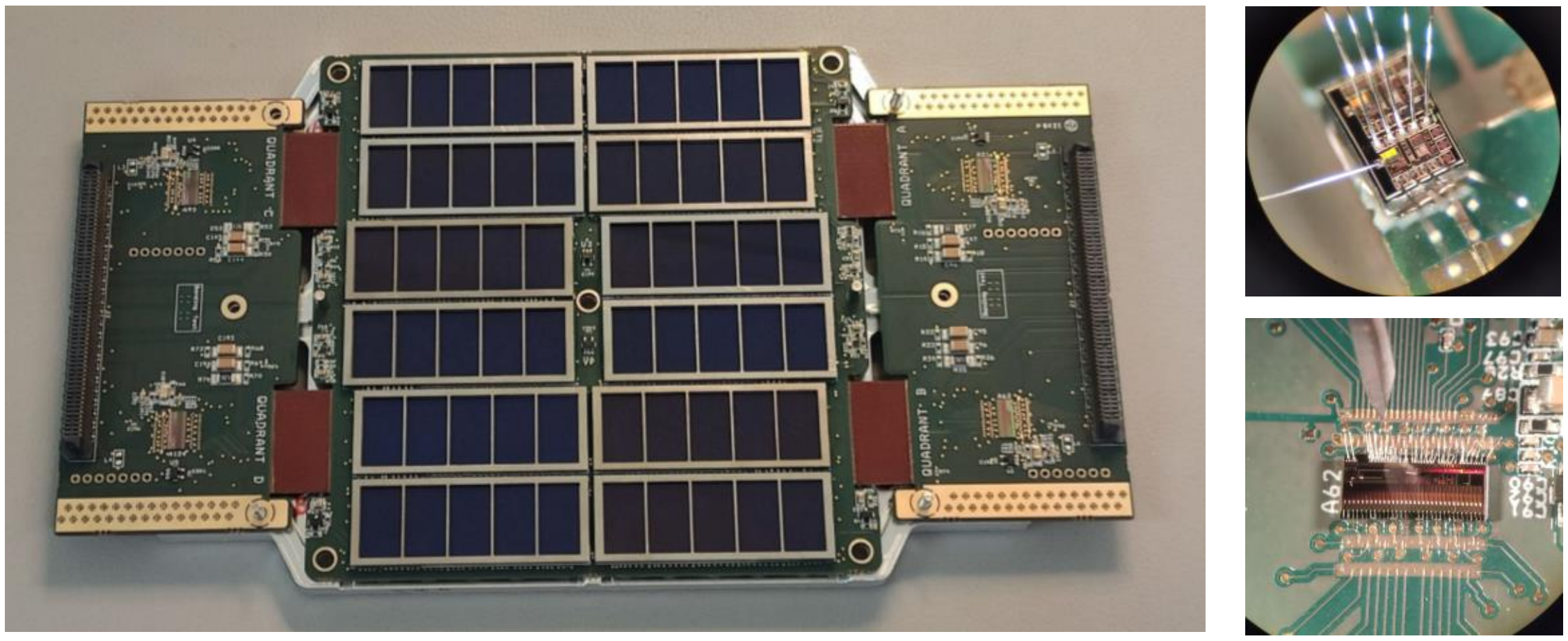

Figure 6 HERMES front-end PCB (left) integrated with 12 SDD arrays, 120 LYRA-FE (top right) and 4 LYRA-BE ASICs (bottom right).

### 3.2 Back-end electronic board (BEE)

The BEE (Figure 7) is the logic block between the front-end ASICs and the Payload Data Handling Unit. The BEE oversees the management of the ASICs configuration, the analog to digital conversion of the ASIC signals and the event time-tagging exploiting the sub-microsecond accuracy of a local chip-scale atomic clock. Moreover, the BEE collects the detector house-keeping data, commands the power lines required by the FEE, manages events data acquisition, transmits science data and HKs to the PDHU as well as to receive and decode tele-commands sent by the PDHU.

The core of the BEE is a SEL immune Intel/Altera Cyclone V FPGA (5CEFA4F23I7N) that implements all the required functions and tasks.

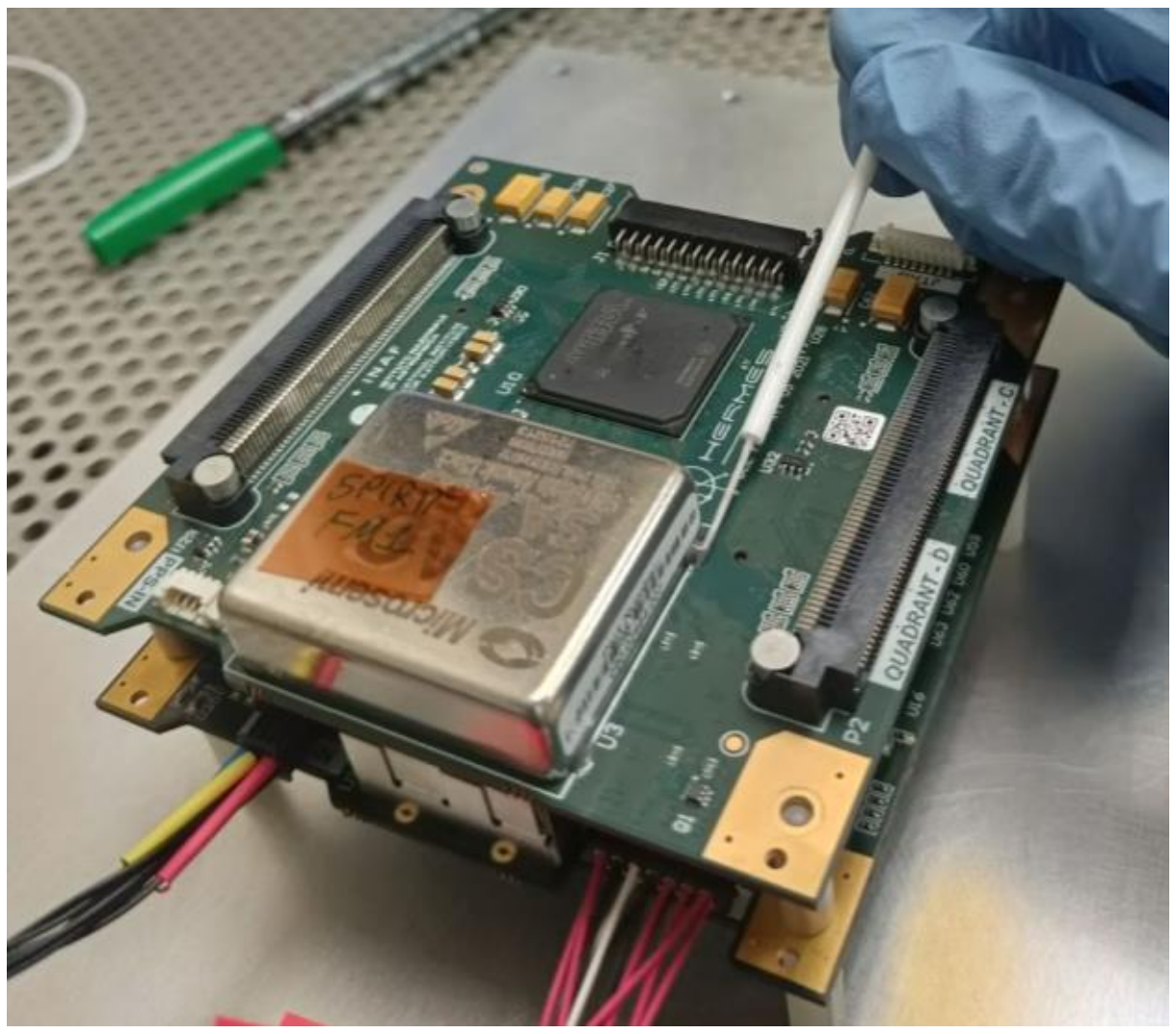

Figure 7 HERMES back-end electronics PCB

The BEE receives, from the LYRA-BE ASICs, the analog event signals produced by the detectors and processed by the front-end board electronics. The signal coming from each LYRA-BE is routed through a preamplifier/buffer to an independent 12 bit, 1 MHz ADC, thus allowing for the simultaneous acquisition of the events detected in different quadrants. With this architecture, the four detector quadrants behave as independent instruments, thus also providing redundancy to the instrument. The TC/TM link between the BEE and the payload data handling unit (PDHU) is ensured through a high speed (5 MHz) SPI interface.

To generate the time reference for events time-tagging a combination of GPS Pulse per Second signal (PPS) from the HERMS service module and a local ultra-stable clock is used. The ultra-stable clock is the space-qualified Microsemi SA.45s chip-scale atomic clock (CSAC), which provides 1 PPS and 10 MHz clock output signals with an accuracy of $\pm 5\times 10^{-11}$, a short-term Allan deviation lower than $10^{-11}$ (1000 s) and a typical aging rate of $9\times 10^{-10}$/month. The BEE FPGA implements 2 counters: one incremented by the CSAC PPS (28 bits) and the other incremented by the 10 MHz clock (24 bits). When the GPS signal is locked, the CSAC PPS signal is synchronized every second to the rising edge of the PPS provided by the satellite bus, thus ensuring a continuous correction for the payload on-board time. In case of GPS not locked, for example when the number of the GPS satellites tracked by the satellite bus GPS receiver is lower than 4 due to the earth occultation, the CSAC still provides the 1 PPS and 10 MHz clock signal in a free-running mode.

Exploiting this architecture, the HERMES Pathfinder timing performance are summarized in Table 2 and Table 3 for the GPS locked and unlocked cases respectively. For the GPS unlocked case (Table 3) an interval of 1500 s from the last GPS lock has been considered.

Table 2 Time precision budget for the GPS locked case

| Mode | Time accuracy (68% c. l.) | Time resolution (68% c. l.) | Total (68% c. l.) |
|---|---|---|---|
| X-mode | 53.4 ns | 320 ns | 324 ns |
| S-mode | 53.4 ns | 216 ns | 222 ns |

Table 3 Time precision budget for the GPS unlocked case

| Mode | Time accuracy (68% c. l.) | Time resolution (68% c. l.) | Total (68% c. l.) |
|---|---|---|---|
| X-mode | 181 ns | 320 ns | 368 ns |
| S-mode | 181 ns | 216 ns | 282 ns |

### 3.3 Power Supply Unit (PSU)

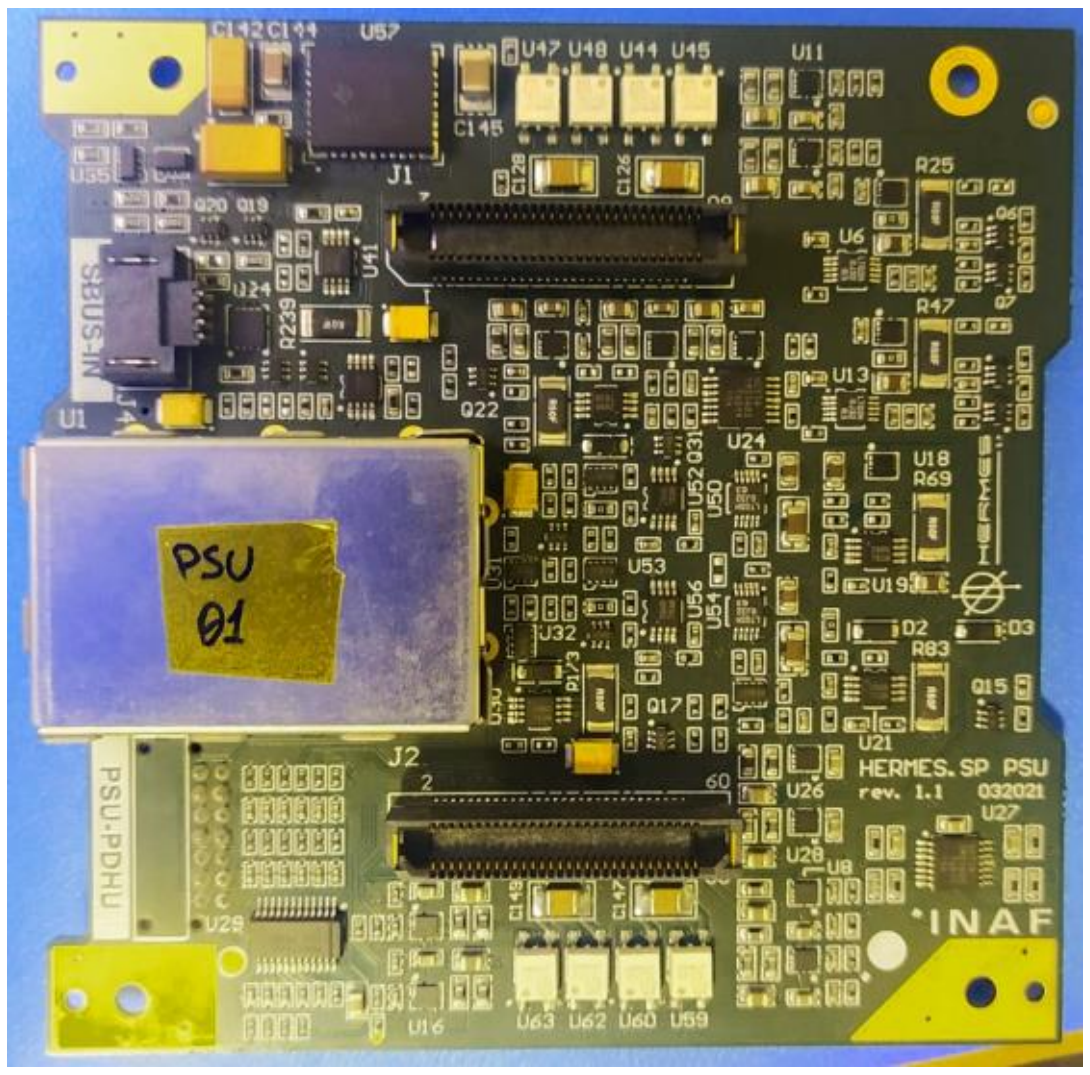

Figure 8 HERMES Pathfinder Power Supply Unit (PSU).

A custom designed Power Supply Unit (PSU, Figure 8) board has been developed and manufactured to provide the power supplies required by the payload. Low voltages needed for the operation of the FEE are generated by ultra-low noise, high PSRR linear Low Drop-Out (LDO) regulators. A DC-DC converter (Picoelectronics 12SAR250) is used for the generation of the high-voltage detector biases while a Texas Instrument LMZ30602 handles the generation of the 1.1V required by the BEE FPGA core.

The PSU receives from the satellite bus four power lines (12 V/80 mA, 5 V/120 mA and 2×3.3 V/1 A). Each line is routed through electronic switches placed on the PSU board and commanded by the PDHU. Another set of switches are placed in series with the main one and commanded by the BEE FPGA firmware.

All the low voltages lines are protected against overcurrent or latch-up in the powered circuitry by means of current monitors which also provide alert lines routed to the BEE FPGA for latch-up monitoring and logging. Special care has been taken in the PSU design to ensure a robust and safe latch-up control. A delay circuit is used to avoid the protection activation due to inrush current, and two of sense amplifiers are implemented in parallel in a redundant configuration for high reliability protection of sensitive power lines. Moreover, all the sense amplifiers are powered downstream the sense resistor. A detailed description of the HERMES PSU can be found in [13].

### 3.4 Payload Data Handling Unit (PDHU)

The HERMES Payload Data Handling Unit (PDHU) is the interface between the spacecraft and the payload. The selected hardware for the PDHU is the Innovative Solutions In Space (ISIS) On-Board computer (iOBC). The iOBC is a flight proven, high performance processing unit based around an ARM9 processor, and offers a multitude of standardized interfaces. Combined with its daughterboard architecture, it allows for easy addition of mission specific electronics or interfaces. A custom-made daughterboard is the current baseline for the PDHU system and provides all the payload-bus electrical interfaces as well as the PHDU internal interface with the other payload subsystems.

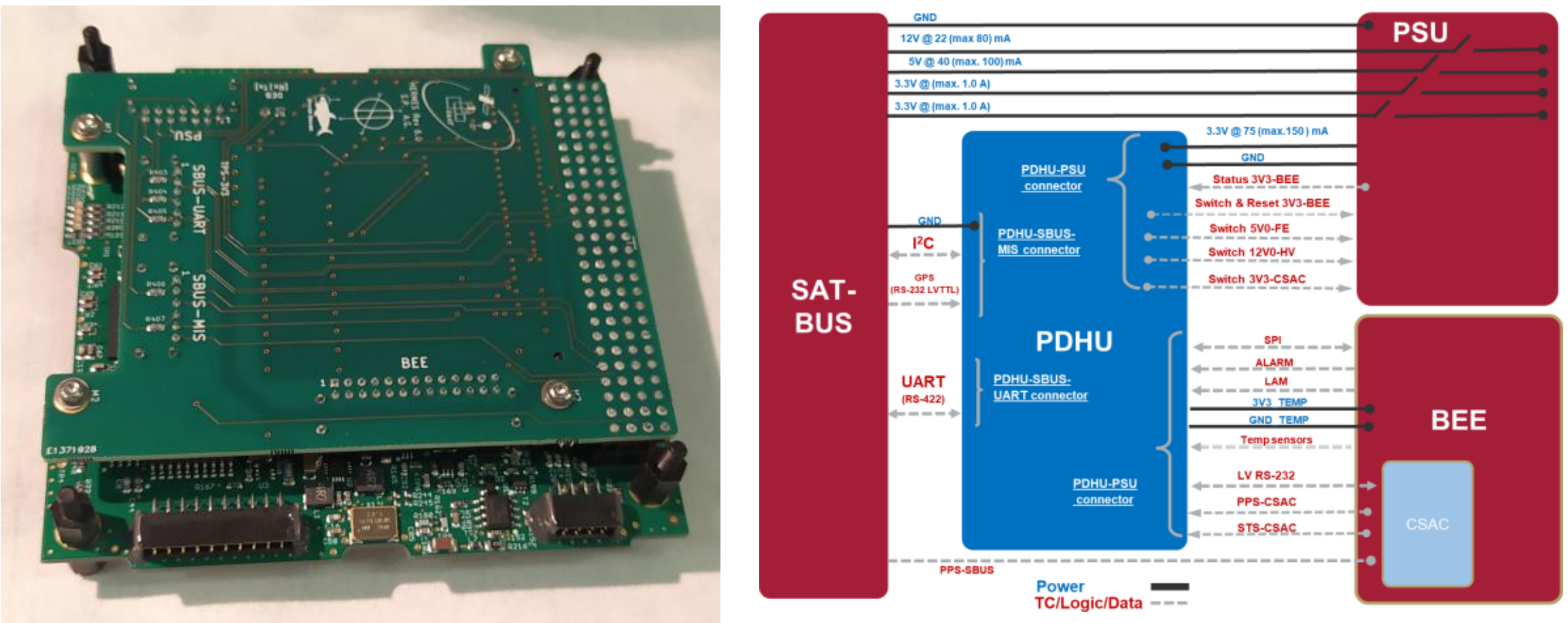

Figure 9 Left: HERMES Payload Data Handling Unit (PDHU) in the motherboard-daughterboard configuration. Right: Schematic view of the P/L electrical interfaces provided by the PDHU

The electrical interfaces between the PDHU and the spacecraft are obtained by two harnesses connected to the PDHU daughterboard (right panel in Figure 9), one providing the cables for the I$^2$C protocol (devoted to direct TCs or error/alarm reporting and power cycling the PDHU) and the LVTTL-RS232 UART (used to receive the GPS data), while the other contains the lines for the UART TM/TC communication on a RS-422 bus.

The PDHU provides the payload central processing unit (CPU) and mass memory, and it oversees interfacing the PSU for voltage line commanding, the BEE for internal TM/TC transmission and CSAC configuration, and the analog temperature sensors of the payload for temperature monitoring and logging. Moreover, the PDHU manages the payload operative modes, generates and filters the photon list, provides the formatting of the scientific and housekeeping data and perform the burst trigger search. A detailed description of the HERMES PDHU and of its functionalities and performance can be found in [14].

### 3.5 Operative modes

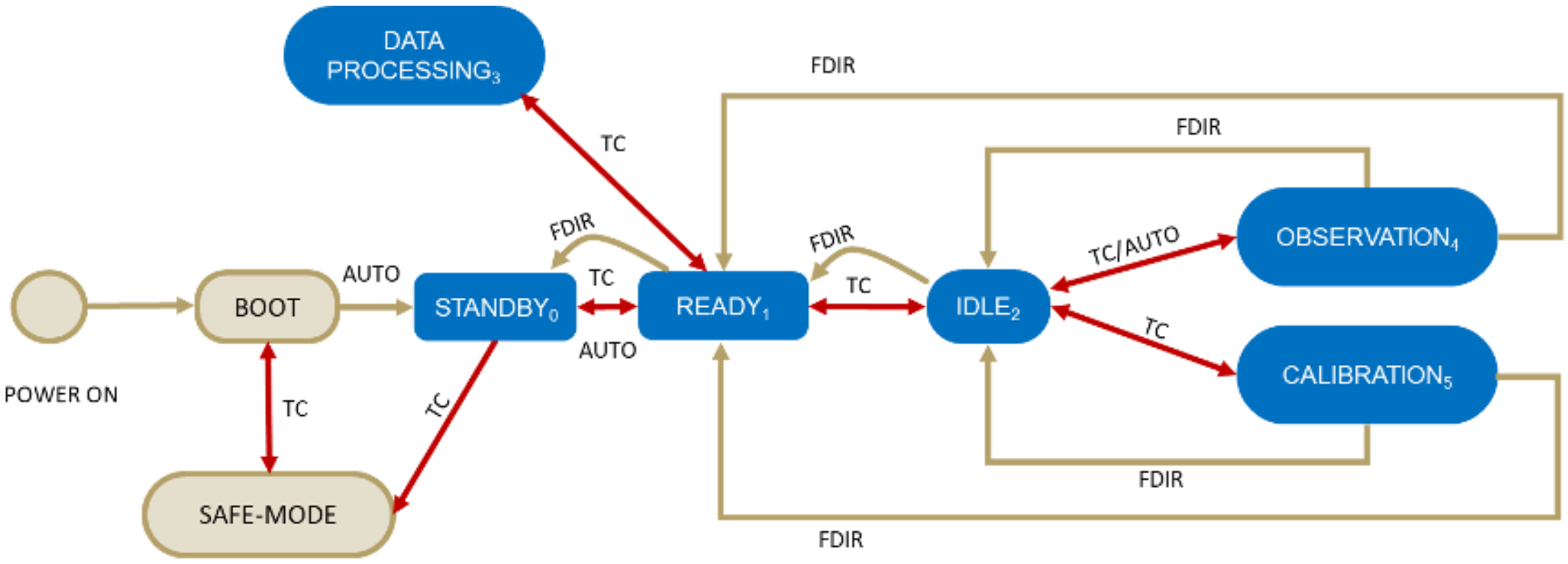

Figure 10 HERMES Payload telecommand (TC) controlled operative modes and Fault Detection, Isolation and Recovery (FDIR) automatic transitions.

- BOOT: Start-up mode at power-on.
- SAFE-MODE: Reserved for exceptional situations as it provides a reduced functionality (mostly only telecommands and diagnostics). It is triggered via either TC or when a major error/malfunction (e.g., latch-up) has been detected.
- STANDBY: After power-on the PDHU moves to STANDBY mode without powering the detector, the PSU and the BEE boards. In this operative mode health checks should be performed and passed, before issuing a TC to move the P/L to the READY state.

- READY: After making health status checks, the PDHU switches ON all power lines with the exclusion of the detector high-voltage bias. All TC/TM interfaces are available as well as HK data. This mode is also used during the S/C SAA passage, activated by a time tagged TC.
- IDLE: The IDLE operative mode provides the same TM/TC interfaces of the READY mode, but with the detector powered-on. Nominally, after observations this is the "go-to" operating mode, although it can also be triggered by the Fault Detection, Isolation and Recovery (FDIR) procedure. Less intensive data processing tasks are also carried out in this mode.
- OBSERVATION AND CALIBRATION: Both instrument-calibration (exploiting the on-board pulse generator) and scientific-acquisition modes can be grouped in a general OBSERVATION mode, whose particular details are configured while in IDLE or STANDBY mode.
- DATA PROCESSING: Its main purpose is to prepare the scientific data packets for transmission to ground. This includes on-board burst-searching algorithms, data cleaning/reducing/compressing, and collection of the housekeeping reports. Although this operational mode is conceived as a stand-alone mode, some of its tasks are commonly run in OBSERVATION, specifically aiming at a prompt acknowledgement of scientifically interesting events.
- FDIR: Although not an operating mode on its own, the Fault Detection Isolation and Recovery (FDIR) approach of the P/L is model-based and relies on a constant monitoring of the on-board sensors (temperatures, currents, and voltages). These readings are compared with an on-board database of updateable nominal ranges. If values are out of the expected ranges, the FDI tasks emits an alert and assess the severity of the fault. Only in high severity cases an automatic power-cycling of the P/L is triggered. This database is kept in the PDHU's FRAM for extra security.

## 4. PAYLOAD PERFORMANCE AND TECHNICAL BUDGETS

In the following, we present the payload performance as experimentally verified with the first HERMES and HERMES-SpIRIT Flight Models (PFM and FM1 respectively).

Figure 11 shows the HERMES PFM (left) and HERMES-SpIRIT FM1 (right) after the completion of the P/L integration procedures. The sole difference between the two payloads is represented by the PCB spacer design, which provide a direct thermal interface with the spacecraft radiator panels in the HERMES 3+3 satellite constellation. At the time of writing the HERMES-SpIRIT FM has been integrated with the SpIRIT mechanical interface and tested in the University of Melbourne laboratories. HERMES PFM is currently (July 2022) undergoing a ground calibration campaign in the IAPS Rome laboratories.

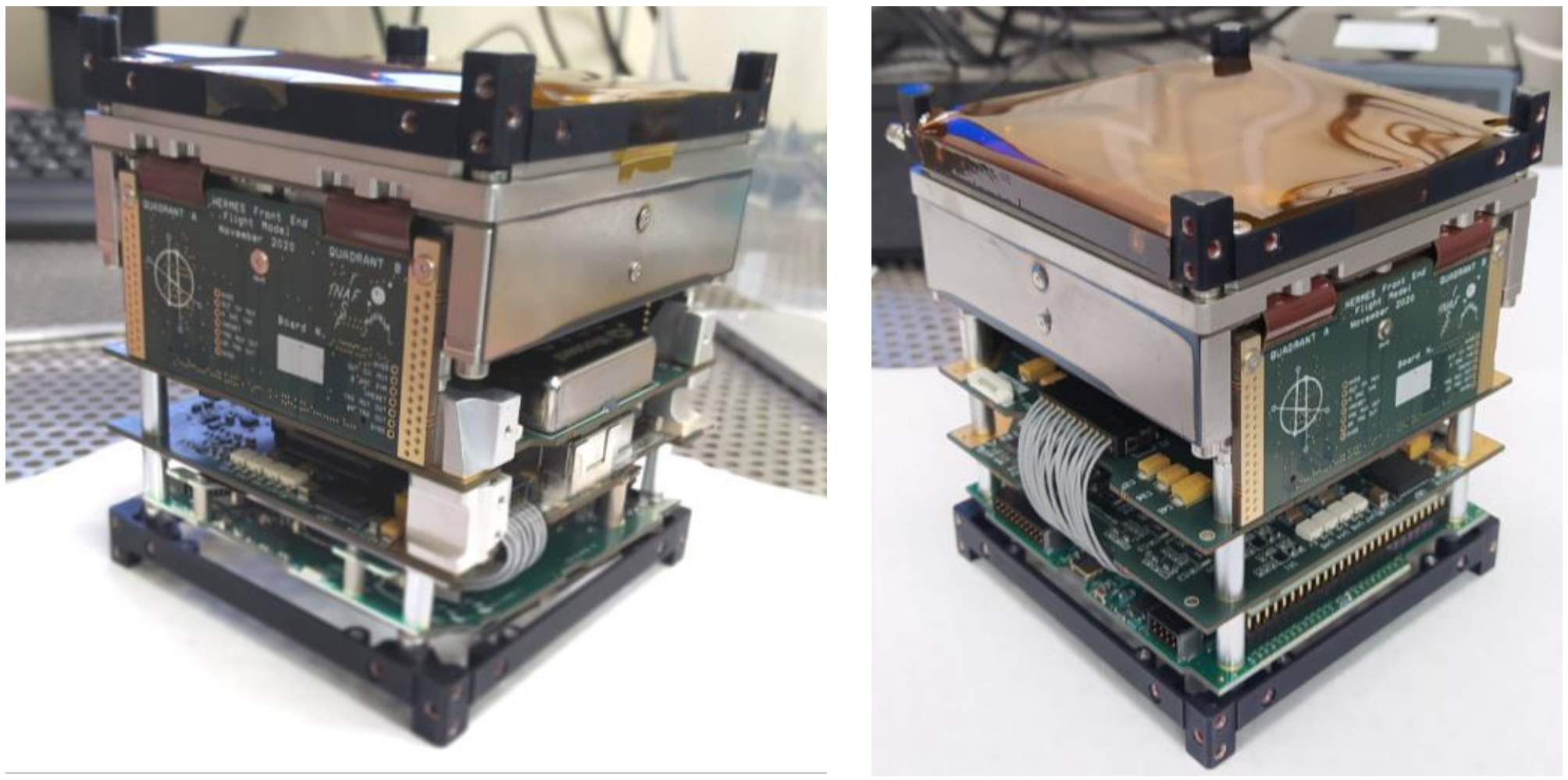

Figure 11 HERMES PFM (left) and HERMES-SpIRIT FM1 (right)

## 4.1 Spectroscopic Performance

Performance characterization and on-ground calibration of the HERMES flight model have been carried out in the INAF-IAPS Rome laboratories in a class 10000 clean room. The flight payload unit has been placed in a climatic chamber with an internal volume of around 1000 $cm^3$ and a series of energy spectra have been acquired between +20 °C and -20 °C by illuminating the detector simultaneously with X-rays ($^{55}Fe$ and $^{109}Cd$) and γ-rays ($^{137}Cs$) sources. Figure 12 shows the calibrated X-mode and S-mode spectra acquired with a detector temperature of -6.5 °C. It is worth noticing that the X-mode lower energy threshold is well below the required 5 keV, with a X-mode energy resolution of ∼300 eV FWHM and a S-mode ΔE/E of about 5%.

A detailed description of the HERMES FM calibration can be found in [15].

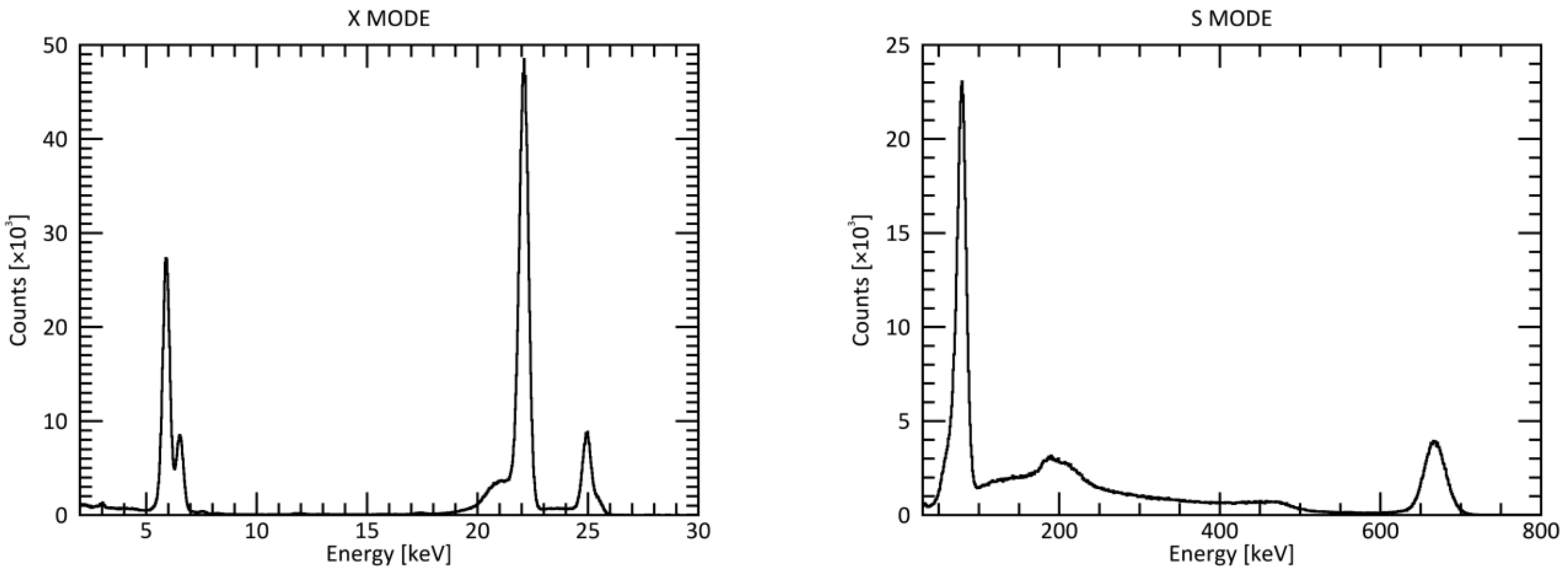

Figure 12 Calibrated energy spectra acquired at –6.5 °C with the HERMES FM1 integrated P/L. Left panel shows the X-mode spectra obtained by illuminating the detector with $^{55}Fe$ and $^{109}Cd$ radioactive sources while the right panel shows the S-mode spectra acquired with $^{109}Cd$ and $^{137}Cs$ radioactive sources.

### 4.2 Power and mass budget

Table 4 reports the FM1 measured mass budget while Table 5 shows the measured P/L power budget in the different operative modes.

Table 4 Measured HERMES P/L mass budget

| Subsystem | Total Mass [g] |
|---|---|
| Detector assembly | 1172.5 |
| PDHU | 97.9 |
| BEE | 98.8 |
| PSU | 72.3 |
| Other items | 37.0 |
| C/S Primary structure | 51.0 |
| P/L Total | **1529.4** |

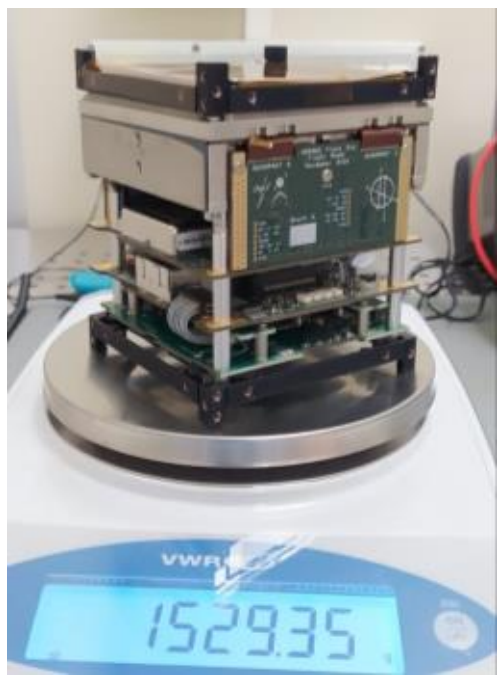

Table 5 HERMES flight model power budget measured for the different operative modes

| OP MODE | SUBSYSTEM | STATUS | POWER [mW] | LINE CURRENTS [mA] | | |
|---|---|---|---|---|---|---|
| | | | | 3.3 | 5.0 | 12.0 |
| BOOT | PDHU MCU | ON | 50 | 15.2 | 0.0 | 0.0 |
| | PDHU | OFF | 0 | | | |
| | PSU | OFF | 0 | | | |
| | BEE | OFF | 0 | | | |
| | FEE TOP | OFF | 0 | | | |
| | FEE SIDES | OFF | 0 | | | |
| | DETECTORS | OFF | 0 | | | |
| | | **TOTAL [mW]** | **50** | | | |
| SAFE | PDHU MCU | ON | 50 | 15.2 | 0.0 | 0.0 |
| | PDHU | OFF | 0 | | | |
| | PSU | OFF | 0 | | | |
| | BEE | OFF | 0 | | | |
| | FEE TOP | OFF | 0 | | | |
| | FEE SIDES | OFF | 0 | | | |
| | DETECTORS | OFF | 0 | | | |
| | | **TOTAL [mW]** | **50** | | | |
| STANDBY | PDHU MCU | ON | 50 | 80.0 | 83.0 | 6.0 |
| | PDHU | ON | 264 | | | |
| | PSU | OFF | 0 | | | |
| | BEE | OFF | 0 | | | |
| | FEE TOP | OFF | 0 | | | |
| | FEE SIDES | OFF | 0 | | | |
| | DETECTORS | OFF | 0 | | | |
| | | **TOTAL [mW]** | **314** | | | |

| OP MODE | SUBSYSTEM | STATUS | POWER [mW] | LINE CURRENTS [mA] | | |
|---|---|---|---|---|---|---|
| | | | | 3.3 | 5.0 | 12.0 |
| READY & DATA PROCESSING | PDHU MCU | ON | 50 | 368.0 | 83.0 | 6.0 |
| | PDHU | ON | 527.5 | | | |
| | PSU | ON | 367 | | | |
| | BEE | ON | 724.4 | | | |
| | FEE TOP | ON | 42.1 | | | |
| | FEE SIDES | ON | 60.4 | | | |
| | DETECTORS | OFF | 0 | | | |
| | | **TOTAL [mW]** | **1771.4** | | | |
| IDLE | PDHU MCU | ON | 50 | 368.0 | 83.0 | 24.0 |
| | PDHU | ON | 527.5 | | | |
| | PSU | ON | 387.6 | | | |
| | BEE | ON | 724.4 | | | |
| | FEE TOP | ON | 42.1 | | | |
| | FEE SIDES | ON | 60.4 | | | |
| | DETECTORS | ON | 125.4 | | | |
| | | **TOTAL [mW]** | **1917.4** | | | |
| OBSERVATION & CALIBRATION | PDHU MCU | ON | 50 | 368.0 | 83.0 | 24.0 |
| | PDHU | ON | 527.5 | | | |
| | PSU | ON | 387.6 | | | |
| | BEE | ON | 724.4 | | | |
| | FEE TOP | ON | 42.1 | | | |
| | FEE SIDES | ON | 60.4 | | | |
| | DETECTORS | ON | 125.4 | | | |
| | | **TOTAL [mW]** | **2139.8** | | | |

## 5. CONCLUSIONS

In this paper we presented the design, integration, and performance characterization of the HERMES pathfinder payload, devoted to the study and fast localization of high energy transients with sub-microsecond time resolution and a wide energy band, spanning from few keV up to 2 MeV. Two payload flight models (PFM and FM1) have been successfully assembled and characterized at the time of writing, while production and test activities of the other five flight modules are currently on-going. Table 6 summarizes the technical and performance budgets of the HERMES pathfinder P/L as measured during the characterization of the PFM and FM1 flight models.

Table 6 Technical and performance budget of the HERMES-TP/SP payload

| **Parameter** | **Value** |
|---|---|
| Payload peak effective area (X-mode & S-mode) | 52 $cm^2$ |
| Field of View | 3.2 sr FWHM |
| Lower energy threshold | ≤ 3 keV |
| Energy resolution X-mode (@ 6 keV) | ≤ 800 eV FWHM |
| Energy resolution S-mode (@ 60 keV) | ≤ 5 keV FWHM |
| Time resolution (68% c.l.) | 320 ns |
| Time accuracy (68% c.l.) | 181 ns |
| Payload mass | 1529 g |
| Payload power in Observation | 2140 mW |
| Payload Telemetry (scientific + engineering) | 930 Mbits/day |

## ACKNOWLEDGMENTS

This work has been carried out in the framework of the HERMES-TP and HERMES-SP collaborations. We acknowledge support from the European Union Horizon 2020 Research and Innovation Framework Programme under grant agreement HERMES-Scientific Pathfinder n. 821896 and from ASI-INAF Accordo Attuativo HERMES Technologic Pathfinder n. 2018-10-H.1-2020.